\documentclass[reprint,twocolumn,superscriptaddress,amsmath,longbibliography,amssymb,prb]{revtex4}
\usepackage{graphicx}% Include figure files
\usepackage{dcolumn}% Align table columns on decimal point
\usepackage{bm}% bold math
\usepackage{multirow}
\usepackage{array}
\usepackage{microtype}
\usepackage{soul}

\usepackage[dvipsnames]{xcolor}
\definecolor{nblue}{rgb}{0.0, 0.0, 1.0}
\definecolor{magenta}{rgb}{0.79, 0.08, 0.48}
\usepackage[colorlinks,linkcolor=nblue,urlcolor=magenta,citecolor=magenta,plainpages=false,pdfpagelabels,breaklinks]{hyperref}

\newcommand{\beq}{\begin{equation}}
\newcommand{\eeq}{\end{equation}}
\newcommand{\bea}{\begin{eqnarray}}
\newcommand{\eea}{\end{eqnarray}}

\begin{document}

\preprint{APS/123-QED}
\title{Pressure-driven superconductivity in the topological insulator GeBi$_4$Te$_7$}

\author{Yalei Huang}
\affiliation{School of Physics, Zhejiang University of Technology, Hangzhou 310023, China}

\author{Na Zuo}
\affiliation{Laboratory of High Pressure Physics and Material Science (HPPMS), School of Physics and Physical Engineering, Qufu Normal University, Qufu 273165, China}

\author{Zheyi Zhang}
\affiliation{School of Physics, Zhejiang University of Technology, Hangzhou 310023, China}

\author{Chunqiang Xu}
\affiliation{School of Physical Science and Technology, Ningbo University, Ningbo 315211, China}

\author{Xiangzhuo Xing}
\email[Electronic address: ]{xzxing@qfnu.edu.cn}
\affiliation{Laboratory of High Pressure Physics and Material Science (HPPMS), School of Physics and Physical Engineering, Qufu Normal University, Qufu 273165, China}

\author{Wen-He Jiao}
\affiliation{School of Physics, Zhejiang University of Technology, Hangzhou 310023, China}

\author{Bin Li}
\email[Electronic address: ]{libin@njupt.edu.cn}
\affiliation{Information Physics Research Center, Nanjing University of Posts and Telecommunications, Nanjing 210023, China}

\author{Wei Zhou}
\email[Electronic address: ]{wei.zhou@cslg.edu.cn}
\affiliation{School of Electronic and Information Engineering, Changshu Institute of Technology, Changshu 215500, China}

\author{Xiaobing Liu}
\affiliation{Laboratory of High Pressure Physics and Material Science (HPPMS), School of Physics and Physical Engineering, Qufu Normal University, Qufu 273165, China}

\author{Dong Qian}
\affiliation
{Key Laboratory of Artificial Structures and Quantum Control (Ministry of Education), School of Physics and Astronomy, Shanghai Jiao Tong University, Shanghai 200240, China}
\affiliation
{Tsung-Dao Lee Institute, Shanghai Jiao Tong University, Shanghai 200240, China}
\affiliation
{Collaborative Innovation Center of Advanced Microstructures, Nanjing 210093, China}

\author{Xiaofeng Xu}
\email[Electronic address: ]{xuxiaofeng@zjut.edu.cn}
\affiliation{School of Physics, Zhejiang University of Technology, Hangzhou 310023, China}

\renewcommand{\thefootnote}{\fnsymbol{footnote}}

\date{\today}

\begin{abstract}
The van der Waals, pseudo-binary chalcogenides ($ACh$)$_m$($Pn_2$$Ch_3$)$_n$ ($A$ = Ge, Mn, Pb, etc.; $Pn$ = Sb or Bi; $Ch$= Te, Se) have recently been reported to host a vast landscape of topological phases of matter, including the quantum anomalous Hall state and topological axion state with quantized magnetoelectric effect. A subgroup in this series, like MnSb$_4$Te$_7$ and GeSb$_4$Te$_7$, can be driven to a superconducting state by applying a physical pressure, making them viable candidates to realize so-called topological superconductivity. However, the role of magnetic fluctuations in this pressure-induced superconductivity remains unclear. Here, we report the pressure-induced multiple superconducting phases in the nonmagnetic GeBi$_4$Te$_7$, accompanied by corresponding structural transitions evidenced from the high-pressure Raman scattering. In comparison with other members in this family, we find the superconducting transition temperature of the nonmagnetic subgroup is significantly higher than their magnetic homologues, possibly hinting at the detrimental role played by the magnetic fluctuations in the superconductivity formation, at least in this pseudo-binary chalcogenide family.
\end{abstract}

\maketitle

Engineering superconductivity in topological materials is generally believed to be an effective approach to achieving topological superconductivity with emergent Majorana fermions, the manipulation of which lays the foundation for future applications in fault-tolerant topological quantum computing~\cite{Hasan-RMP,Nayak-RMP,armitage-review,SC-Zhang-review}. Among the various means of tuning, such as chemical substitution or intercalation, pressure proves to be a clean and fruitful way in the quest for topological superconductor candidates. Indeed, superconductivity achieved in this manner has been observed in some archetypal topological insulator (TI) candidates. For example, TIs Bi$_2$Se$_3$ and Sb$_2$Te$_3$ become bulk superconductors by the application of pressure, with a maximum $T_c$ value of 7 K and 6.3 K, respectively~\cite{Bi2Se3-PRL,ChangqingJin-SR}.

\begin{figure*}
\begin{center}
\includegraphics[width=\textwidth]{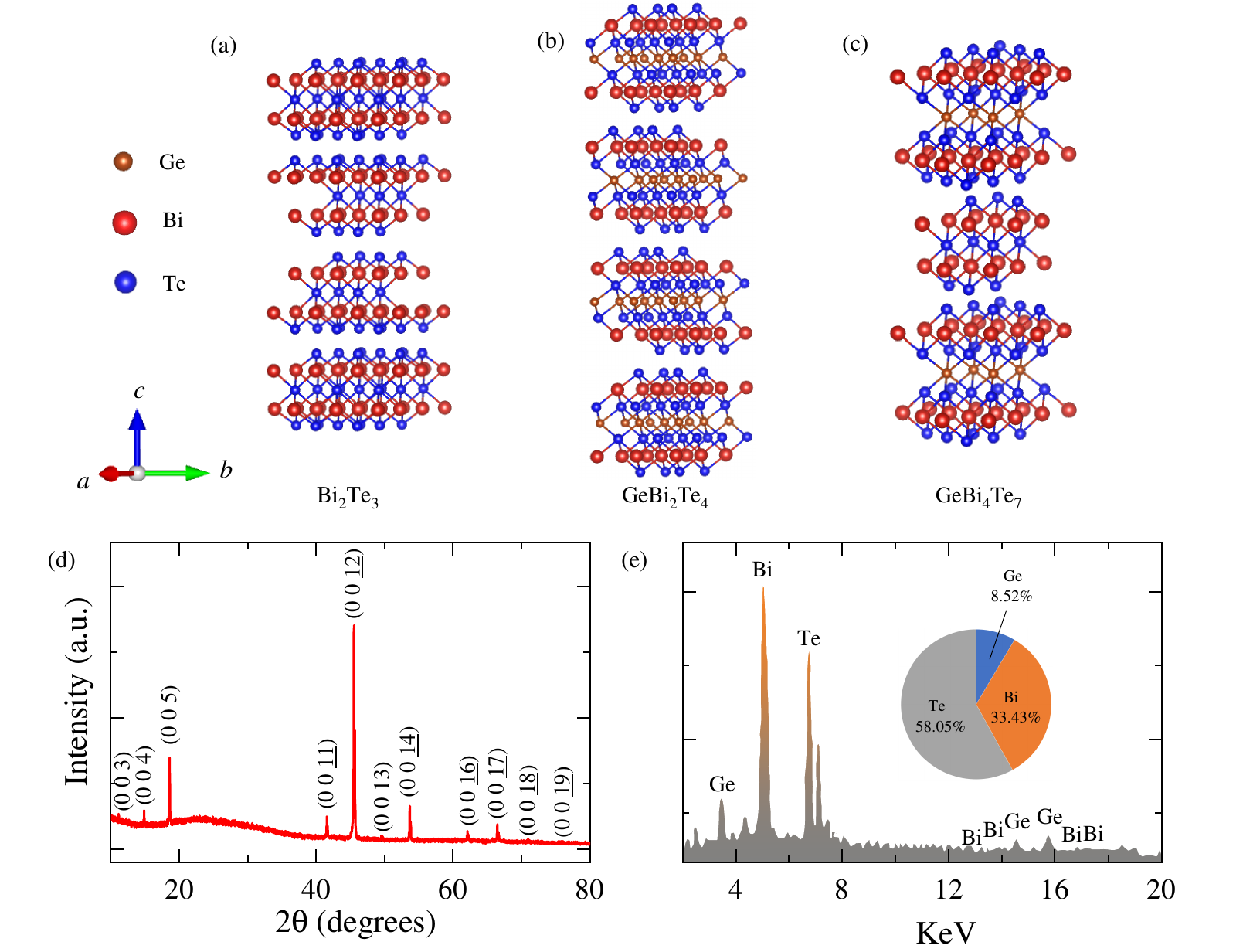}
\caption{(a)-(c) The crystal structure of the pseudo-binary chalcogenides ($ACh$)$_m$($Pn_2$$Ch_3$)$_n$, represented by $A$=Ge, $Pn$=Bi, $Ch$=Te. In panel (a), $m$=0, $n$=1. In panel (b), $m$=1, $n$=1. In panel (c), $m$=1, $n$=2. (d) The single crystal XRD of the GeBi$_4$Te$_7$ sample. Only (00$l$) peaks have been observed, indicating the sample top facet is along the $c$ axis. (e) A typical EDX pattern showing the stoichiometric ratio of Ge:Bi:Te is very close to 1:4:7.  } \label{Fig1}
\end{center}
\end{figure*}

Among a large number of as yet discovered topological materials, the multilayered, pseudo-binary chalcogenides ($ACh$)$_m$($Pn_2$$Ch_3$)$_n$ ($A$ = Ge, Mn, Pb, etc.; $Pn$ = Sb or Bi; $Ch$= Te, Se) are of particular interest~\cite{PRB-GeBiSbTe,PRB-GeSbTe-Zhou,Qi-PRM,Qi-CPL,GuoYF-PRL}. On one hand, $Pn_2Ch_3$, i.e., $m$=0, $n$=1, were the first validated bulk TIs, thereafter opening the door to search for versatile topological properties of hundreds of other materials. Structurally, these Bi$_2$Te$_3$-type compounds crystallize in the tetradymite-like layered structure, consisting of Te-Bi-Te-Bi-Te quintuple layers (5L,QLs) that are stacked along the $c$-axis by van der Waals forces (see Fig.~\ref{Fig1}(a) for the structural motif)~\cite{NC-PbBi4Te7}. On the other hand, the $ACh$ bilayer can readily intercalate the QLs, forming the $Ch$-$Pn$-$Ch$-$A$-$Ch$-$Pn$-$Ch$ septuple layers (7L,SLs) (Fig.~\ref{Fig1}(b)). With these QLs and SLs as the basic building elements, new van der Waals compounds ($ACh$)$_m$($Pn_2Ch_3$)$_n$ can be formed. For concreteness, in the 124 phase, i.e., $m$=1, $n$=1, the structure is established by the stacking of SLs only (Fig.~\ref{Fig1}(b)), whereas in the 147 phase, i.e., $m$=1, $n$=2, it consists of alternating stacking of QLs and SLs along the $c$-axis, forming a natural superlattice (Fig.~\ref{Fig1}(c)). Remarkably, most members in this family were reported to be 3D TIs, offering a tunable platform to observe diverse topological phenomena, including topological superconductivity.

\begin{figure*}
\begin{center}
\includegraphics[width=\textwidth]{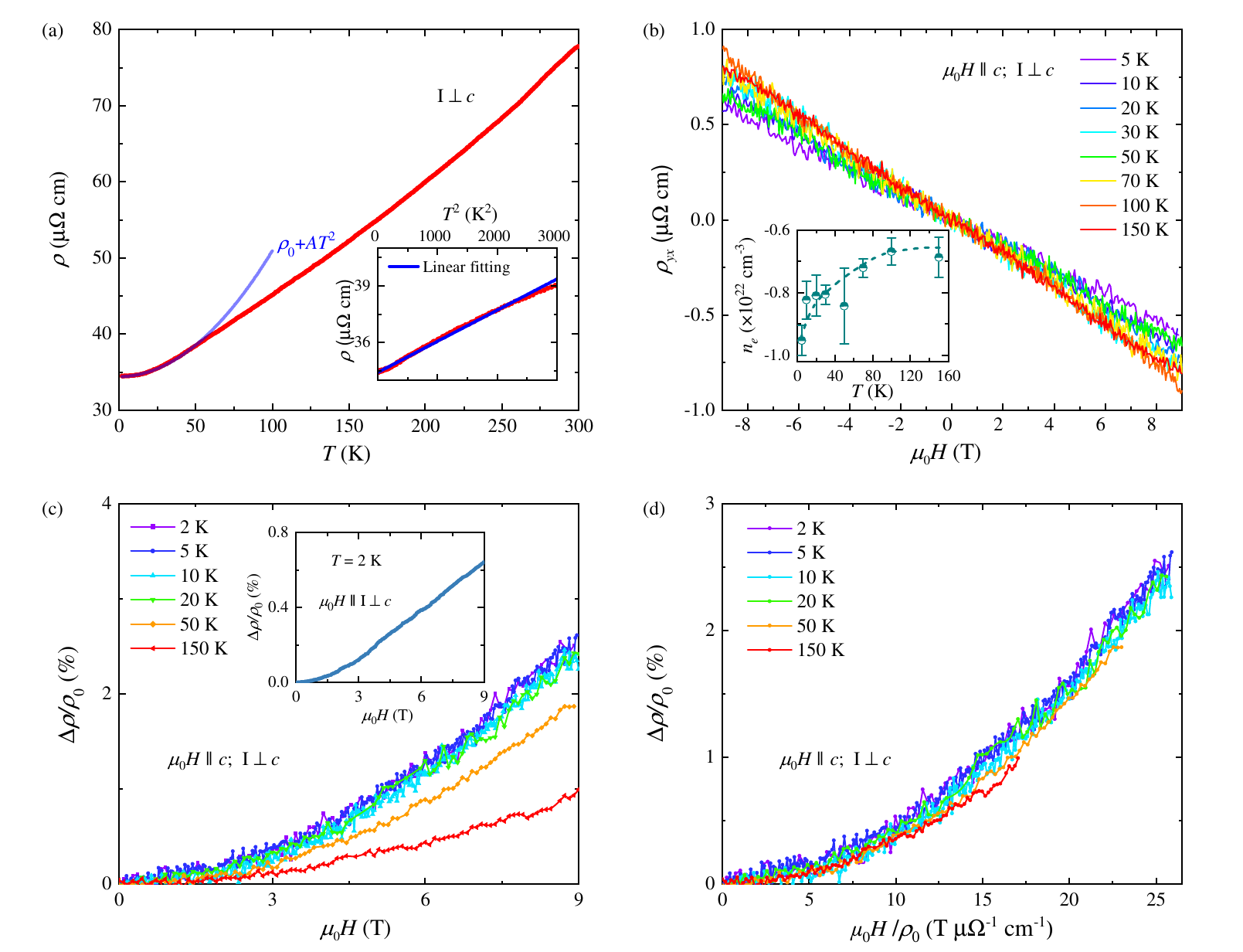}
\caption{(a) The zero-field in-plane resistivity. Below $\sim$50 K, a Fermi liquid behavior is seen which is better resolved in the inset when plotted as $\rho$ vs $T^2$. (b) The Hall resistivity $\rho_{yx}$ at some fixed temperatures showing quasi-linear scaling with field. The inset shows the electron density $n_e$ extracted from the single-band fitting. (c) The TMR at some fixed temperatures. The inset shows the longitudinal magnetoresistance at 2 K as a comparison. (d) The Kohler's plot of the data in panel (c). } \label{Fig2}
\end{center}
\end{figure*}

In the past few years, substantial research efforts have been devoted to tuning the superconductivity in these pseudo-binary chalcogenides ($ACh$)$_m$($Pn_2$$Ch_3$)$_n$~\cite{Qi-PRM,PRB-GeSbTe-Zhou,PRB-GeSb2Te4,PRB-GeSb2Te4-amorphous}. For instance, besides the above mentioned Bi$_2$Se$_3$ and Sb$_2$Te$_3$, superconductivity was reported in SnSb$_2$Te$_4$ under pressure where $T_c$ is gradually enhanced with increasing pressure up to 33 GPa~\cite{SnSb2Te4}. Superconductivity was later revealed in both crystalline and amorphous forms of GeSb$_2$Te$_4$, with two superconducting phases being observed for $P <$ 40 GPa~\cite{PRB-GeSb2Te4,PRB-GeSb2Te4-amorphous}. More recently, superconductivity was also found in the pressurized GeSb$_4$Te$_7$~\cite{PRB-GeSbTe-Zhou}.

Surprisingly though, superconductivity was not observed in the pressurized magnetic MnBi$_2$Te$_4$ and MnBi$_4$Te$_7$~\cite{Qi-CPL}. Intriguingly, MnSb$_4$Te$_7$ was recently reported to show a domelike phase diagram with a maximum $T_c$ of only 2.2 K at $\sim$50 GPa~\cite{Qi-PRM}, significantly lower than its nonmagnetic sibling GeSb$_4$Te$_7$ which has a maximum $T_c$ of approximate 8 K at 35 GPa~\cite{PRB-GeSbTe-Zhou}, leading to the conjecture that the magnetic fluctuations in the Mn-based homologues may be adverse to the formation of Cooper pairs.

In this study, we report the pressure-induced multiple superconducting phases in GeBi$_4$Te$_7$. To be specific, we observed three superconducting phases between 8 GPa and 42 GPa, with the maximum $T_c$ over 8 K at $\sim$30 GPa. These three superconducting phases are driven by the putative structural phase transitions, in analogy to what was observed in GeSb$_4$Te$_7$~\cite{PRB-GeSbTe-Zhou}. By comparing with other members in this class of materials, in particular the magnetic MnBi$_4$Te$_7$~\cite{Qi-CPL}, our study suggests that the magnetic fluctuations play a possible unfavorable role in the superconductivity formation, at least in these pseudo-binary chalcogenide systems.

\begin{figure*}
\begin{center}
\includegraphics[width=\textwidth]{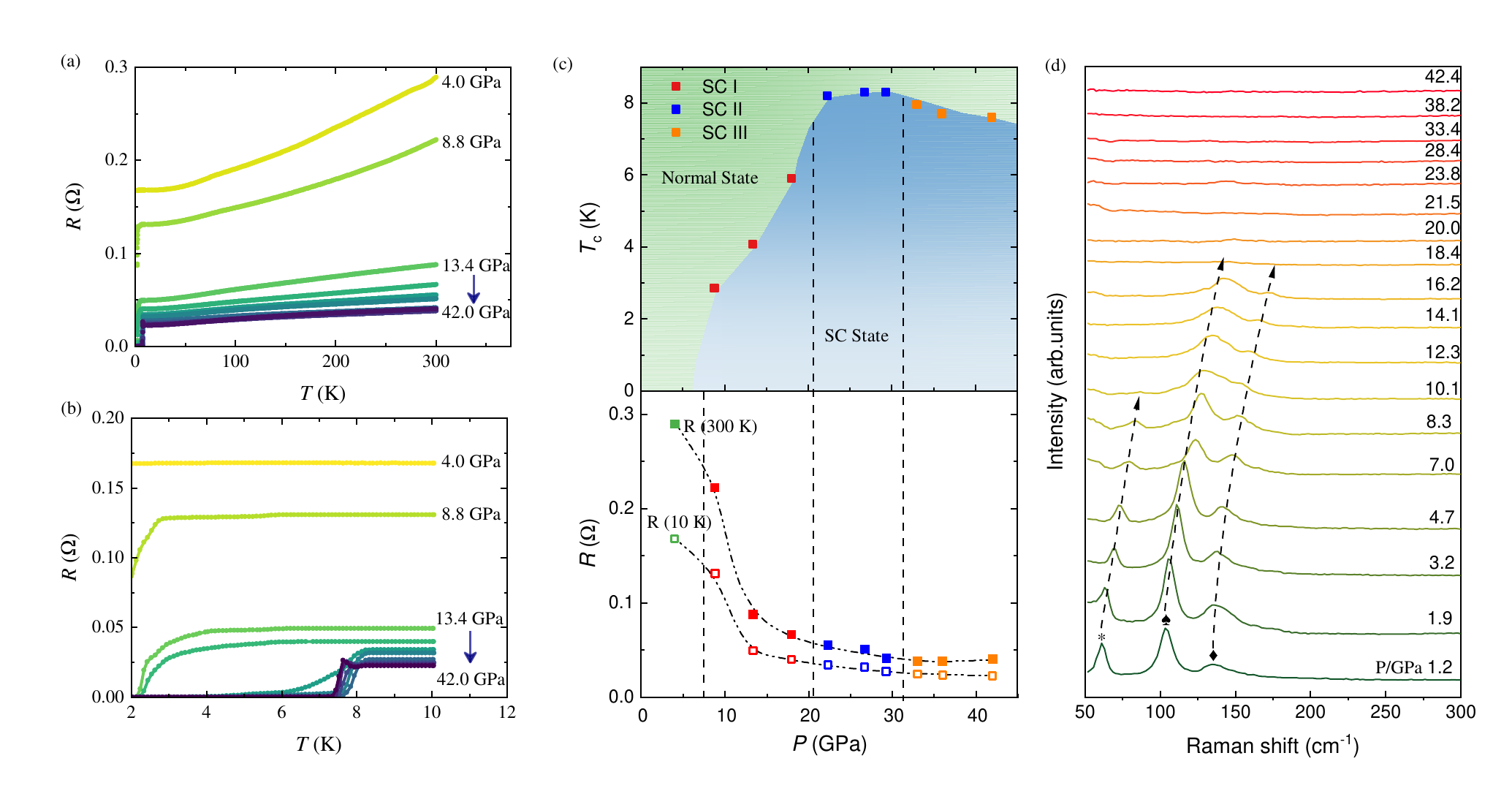}
\caption{(a) The resistance down to 2 K at various pressures. (b) The zoomed-in view of the low-temperature resistance. (c) The pressure phase diagram (top) and the resistances at 300 K and 10 K (bottom). (d) The pressure evolution of the Raman spectra at different pressures. } \label{Fig3}
\end{center}
\end{figure*}

Single crystals of GeBi$_4$Te$_7$ were synthesized by the self-flux method~\cite{Cava}. The elements of Ge, Bi and Te were mixed in a molar ratio of 1:4:15, placed into an alumina crucible, and sealed in an evacuated quartz tube. Subsequently, the mixture was heated to 1050 $^\circ$C over a period of 12 hours, then slowly cooled down to 500 $^\circ$C at a rate of 2 $^\circ$C per hour, followed by centrifugation to remove the excess flux. The van der Waals crystals of GeBi$_4$Te$_7$, with a typical size of 2$\times$2$\times$0.1 mm$^3$, were achieved. The composition of the single crystals was determined by the energy-dispersive x-ray spectroscopy (EDS) on a scanning electron microscope (Hitachi S-3700N) equipped with an Oxford Instruments X-Max spectrometer. X-ray diffraction (XRD) was carried out on a PANalytical x-ray diffracometer (Model EMPYREAN) with monochromatic Cu $K$$\alpha$ radiation at room temperature. The electrical transport properties were measured in a cryostat from Cryogenic Instruments using a standard four-probe method. The MR and Hall resistivity were measured by changing the polarities of the field and symmetrizing and anti-symmetrizing the data, respectively. High-pressure transport experiments were conducted in a nonmagnetic Be-Cu diamond anvil cell with NaCl as the pressure transmitting medium. A single crystal with dimensions of 130$\times$50$\times$10 $\mu$m$^3$ was loaded together with some ruby powder. High-pressure Raman spectroscopy data were collected by a Renishaw inVia Raman system with a laser wavelength of 532 nm.

The electronic structure of GeBi$_4$Te$_7$ was investigated using first-principles density functional theory (DFT) calculations implemented within the WIEN2K package~\cite{Wien2k}, employing the full-potential linearized augmented plane wave (FLAPW) method. We utilized the Wu-Cohen generalized gradient approximation (GGA)~\cite{GGA} for the exchange-correlation functional, with 18$\times$18$\times$2 $k$-point mesh in the self-consistent field calculation. To ensure a comprehensive and accurate description of the electronic structure, the spin-orbit coupling (SOC) were incorporated in all calculations. The Fermi surface was mapped using a high-density 40$\times$40$\times$7 $k$-point mesh across the complete Brillouin zone (BZ). We checked the topological properties using the Elementary Band Representations (EBRs) method~\cite{ebr1, ebr2}. At each maximal $k$-vector, we first determined the collection of irreducible representations. Subsequently, by applying compatibility relations and referencing the EBRs, we verified whether the band structure can be decomposed into linear combinations of EBRs with integer coefficients. We also constructed a tight-binding model using maximally localized Wannier functions (MLWFs)~\cite{Mostofi2014An} as implemented in the Wannier Tools package~\cite{Wu2017WannierTools}. The tight-binding Hamiltonian was built using 72 bands, incorporating Ge 4$p$, Bi 6$p$ and Te 5$p$ orbitals as the basis for the topological analysis.

The single crystal XRD pattern at ambient conditions has been demonstrated in Fig.~\ref{Fig1}(d). All peaks can be well indexed by the (00$l$) reflections of the trigonal $P\bar{3}m1$ (No. 164) space group, indicating the crystallographic $c$-axis is normal to the sample top facet. The calculated $c$-axis lattice constant is $c$=23.90 $\texttt{{\AA}}$, in excellent agreement with the published value~\cite{Cava}. The chemical composition of the single crystals was checked by the EDS with a typical spectrum given in Fig.~\ref{Fig1}(e). The stoichiometry of the constituent Ge, Bi and Te was seen to be very close to 1:4:7 from the multiple-spot scanning on the surface, confirming the high quality of the GeBi$_4$Te$_7$ single crystals used in this study.

The in-plane resistivity (I$\parallel$$ab$) has been measured from room temperature down to 2 K (Fig.~\ref{Fig2}(a)). The metallic $\rho$($T$) profile is seen, with the residual resistivity ratio (RRR) of $\sim$2, indicating the poor metallicity. Below $\sim$50 K, resistivity follows a $T^2$ behavior, indicating a Fermi liquid ground state~\cite{Xiafeng-thesis}. This $T^2$ scaling is better visualized in the inset of Fig.~\ref{Fig2}(a). The Hall resistivity $\rho_{yx}$ has been measured at some fixed temperatures between 5 K and 150 K, as shown in Fig.~\ref{Fig2}(b). It is seen that $\rho_{yx}$ has a negative slope and is quasi-linear in field, suggesting the dominant electron-type carriers. The carrier density $n_e$ can be extracted from $n_e=\frac{1}{eR_H}$, where $R_H$=$\frac{\rho_{yx}}{B}$. The carrier density $n_e$, as a function of temperature, is displayed in Fig.~\ref{Fig2}(b) inset. With decreasing temperature, $n_e$ increases slightly.

The transverse magnetoresistance (TMR), with the current applied along the $ab$ plane and the magnetic field along the $c$-axis, has been measured, as plotted in Fig.~\ref{Fig2}(c). As noted, the TMR is quadratic in field and its magnitude is rather small, with $\Delta\rho/\rho_0$ of only 2.6\% at 2 K and 9 T. The longitudinal magnetoresistance is a factor of 4 smaller, as shown in the inset of Fig.~\ref{Fig2}(c). The Kohler's scaling, plotted as $\Delta\rho/\rho_0$ as a function of $\mu_0 H/\rho_0$ at different temperatures as shown in Fig.~\ref{Fig2}(d), is well obeyed in the measured temperature range, suggesting no exotic electron scattering or phase transitions~\cite{XiaoZL-PRX,Hussey-Science}. It is intriguing to examine the Kohler's rule in the magnetic analogues, such as MnBi$_2$Te$_4$ and MnBi$_4$Te$_7$. However, there is no such reports in the literature thus far.

We now turn to the pressure effects on the transport properties of GeBi$_4$Te$_7$. As shown in Fig.~\ref{Fig3}(a), $R$($T$) curves have been measured under pressures up to 42 GPa. The pressure is seen to overall suppress the resistance in the whole temperature range. The low-$T$ region, as enlarged in Fig.~\ref{Fig3}(b), signifies the pressure-induced superconductivity. Specifically, at $\sim$8.8 GPa, superconductivity starts to set in below $\sim$2.8 K. Further increase of pressure boosts $T_c$ to above 8 K when $P$$>$ 22 GPa. Above 30 GPa, however, $T_c$ begins to decrease with increasing pressure. The resultant phase diagram is illustrated in Fig.~\ref{Fig3}(c), highlighting the pressure-induced superconducting phase in this multilayered pseudo-binary chalcogenide GeBi$_4$Te$_7$. The resistances at 300 K and at 10 K, are extracted in Fig.~\ref{Fig3}(c) as a function of pressure. Reminiscent of GeSb$_4$Te$_7$~\cite{PRB-GeSbTe-Zhou}, four regions are tentatively compartmentalized based on the pressure dependent $T_c$ and $R$(300 K) and $R$(10 K) (see Fig.~\ref{Fig3}(c)).

To have more insights on the possible pressure-induced structural transitions in GeBi$_4$Te$_7$, we performed the Raman scattering measurements under pressure, as shown in Fig.~\ref{Fig3}(d). At 1.2 GPa, there are three Raman modes at 62 cm$^{-1}$, 103 cm$^{-1}$ and 135 cm$^{-1}$, respectively. With increasing pressure, these modes shift to higher wave numbers. Above 8.3 GPa, the first Raman peak becomes unresolvable while the other two remain to move to higher wave numbers with increasing pressure, possibly indicating a structural transition. Above 20 GPa, no peaks can be observed, indicating another possible structural transition or pressure-induced amorphization.

\begin{figure*}
\begin{center}
\includegraphics[width=14cm]{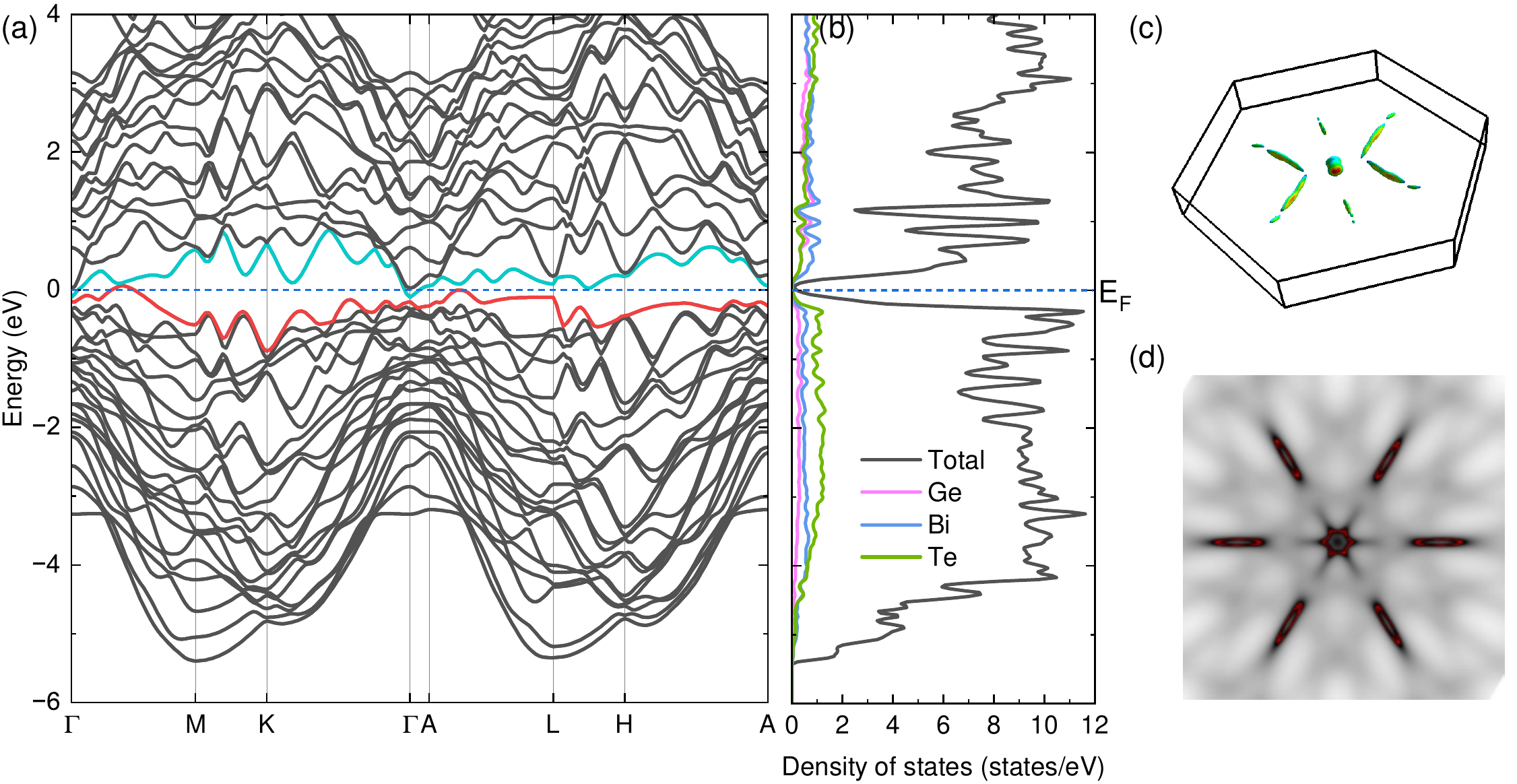}
\caption{Electronic properties of GeBi$_4$Te$_7$: (a) Band structure along the high-symmetry path $\Gamma$--X--M--$\Gamma$--A--L--H--A in the Brillouin zone, highlighting the two bands (colored) that cross the Fermi level. (b) Total and partial density of states (DOS). (c) Three-dimensional visualization of the Fermi surfaces  with color mapping representing the Fermi velocity. (d) Top view of the Fermi surfaces obtained from Wannier function calculations, illustrating the six-fold symmetry.}
\label{Fig4}
\end{center}
\end{figure*}

Figure~\ref{Fig4} shows that with spin-orbit coupling, the electronic band structure exhibits two splitting bands crossing the Fermi level. The three-dimensional and top view of the Fermi surface reveals six-fold symmetry, consisting of one electron pocket at the BZ center and six groups of multi-component hole pockets distributed around the BZ. We analyzed the topological properties of GeBi$_4$Te$_7$ using topological indices calculated from the EBRs method. The set of bands below the Fermi level cannot be expressed as a linear combination of EBRs alone but can be represented as a linear combination of EBRs and disconnected parts of EBRs. Our analysis identifies the compound to be a topological insulator with topological indices z$_{2w, 3}$=1 and z$_4$=1.

\begin{table*}
	%\color{blue}	
	 	\caption{The pressure-induced superconductivity in the pseudo-binary chalcogenides ($ACh$)$_m$($Pn_2$$Ch_3$)$_n$ ($A$ = Ge, Mn, Pb, etc.; $Pn$ = Sb or Bi; $Ch$= Te, Se). Here '$\times$' denotes that no superconductivity is observed under pressure. "?" means no report on the pressure study in the literature. $T_c$ here indicates the maximum $T_c$ in the pressure range studied, i.e., usually up to $\geq$50 GPa.}
	\begin{ruledtabular}
		\begin{tabular}{cccccc}
			magnetic $$ & SC under pressure $$ &\multicolumn{2}{c}{nonmagnetic}&\multicolumn{2}{c}{SC under pressure}\\
			\hline
			MnBi$_2$Te$_4$ & $\times$~\cite{Qi-CPL} &\multicolumn{2}{c}{GeBi$_2$Te$_4$}&\multicolumn{2}{c}{$T_c$$\sim$8 K}~\cite{Cui-GeBi2Te4}\\
			MnBi$_4$Te$_7$ & $\times$~\cite{Qi-CPL} &\multicolumn{2}{c}{GeBi$_4$Te$_7$}&\multicolumn{2}{c}{$T_c$$\sim$8.3 K} [this work]\\
            MnSb$_2$Te$_4$ & $\times$~\cite{MnSb2Te4-pressure} &\multicolumn{2}{c}{GeSb$_2$Te$_4$}&\multicolumn{2}{c}{$T_c$$\sim$7 K}~\cite{PRB-GeSb2Te4,PRB-GeSb2Te4-amorphous}\\
            MnSb$_4$Te$_7$ & $T_c$$\sim$2 K~\cite{Qi-PRM} &\multicolumn{2}{c}{GeSb$_4$Te$_7$}&\multicolumn{2}{c}{$T_c$$\sim$8 K}~\cite{PRB-GeSbTe-Zhou}\\
            MnBi$_6$Te$_{10}$ & $\times$~\cite{MnBi6Te10-pressure} &\multicolumn{2}{c}{SnSb$_2$Te$_4$}&\multicolumn{2}{c}{$T_c$$\sim$7.4 K}~\cite{SnSb2Te4}\\
            MnSb$_6$Te$_{10}$ & $?$ &\multicolumn{2}{c}{Bi$_2$Te$_3$}&\multicolumn{2}{c}{$T_c$$\sim$8 K}~\cite{Bi2Te3-pressure-Jinchangqing}\\
             &  &\multicolumn{2}{c}{Sb$_2$Te$_3$}&\multicolumn{2}{c}{$T_c$$\sim$6.3 K}~\cite{ChangqingJin-SR}\\
             &  &\multicolumn{2}{c}{Bi$_2$Se$_3$}&\multicolumn{2}{c}{$T_c$$\sim$7 K}~\cite{Bi2Se3-PRL}\\
		\end{tabular}
	\end{ruledtabular}
\end{table*}

Before closing, let us compare the high-pressure phase diagram of GeBi$_4$Te$_7$ extracted from this study, with those of other members in this class of pseudo-binary chalcogenides whose pressure effects have been investigated in the literature. The most direct comparison is with the isostructural MnBi$_4$Te$_7$ that has an antiferromagnetic ground state~\cite{Qi-CPL}. Pressure changes dramatically the resistivity of MnBi$_4$Te$_7$ and induces a non-monotonic evolution of $\rho$($T$) due to two high-pressure phase transitions~\cite{Qi-CPL}. However, no superconductivity is observed up to $\sim$50 GPa. Applying pressure on MnBi$_2$Te$_4$ induces metal-semiconductor-metal transitions with possible structural transitions and the concomitant changes in the topological properties~\cite{Qi-CPL}. Again, no superconductivity develops up to $\sim$50 GPa. Pressure-induced superconductivity was not reported in the magnetic subgroup of this pseudo-binary chalcogenide family until the very recent observation of superconductivity under pressure in the magnetic MnSb$_4$Te$_7$~\cite{Qi-PRM}. Compared with MnBi$_2$Te$_4$ and MnBi$_4$Te$_7$, however, the magnetic interaction seems to be weaker in MnSb$_4$Te$_7$~\cite{Qi-PRM}. Moreover, compared with the nonmagnetic GeSb$_4$Te$_7$, superconducting $T_c$ is much lower in the magnetic MnSb$_4$Te$_7$; superconducting $T_c$ was observed when $P$ $>$ 30 GPa, reaching a maximum $T_c$ of 2 K at 50 GPa in MnSb$_4$Te$_7$ while in GeSb$_4$Te$_7$, $T_c$ sets in above 10 GPa and continues to rise with increasing pressure with no sign of saturation up to $\sim$35 GPa~\cite{Qi-PRM,PRB-GeSbTe-Zhou}. By the same token, the nonmagnetic GeBi$_2$Te$_4$ has a $T_c$ over 8 K at the pressure of $\sim$15 GPa~\cite{Cui-GeBi2Te4}, while no superconductivity was reported for the pressurized MnBi$_2$Te$_4$~\cite{Qi-CPL}. These observations do seem to suggest that magnetic fluctuations play an adverse role in the superconductivity formation, at least in this class of materials. This argument is further buttressed up by the reported superconductivity in GeSb$_2$Te$_4$ under pressure ($T_c$=7 K at $P$=20 GPa) that is nonmagnetic in origin~\cite{PRB-GeSb2Te4,PRB-GeSb2Te4-amorphous}. Table I compares the pressure effects on superconductivity in the the magnetic and nonmagnetic pseudo-binary chalcogenides that have been studied thus far.

In summary, we have successfully grown the single crystals of the layered van der Waals pseudo-binary chalcogenide GeBi$_4$Te$_7$. We revealed the (multiple) superconducting phases under pressures up to 42 GPa. Its topological properties have also been analyzed by the first-principles calculations. By comparing with the pressure effects in other members of this class of pseudo-binary chalcogenides, our study suggests that the magnetic fluctuations may be unfavorable to the formation of superconductivity in this family of topological materials.

%\bibliography{GeBi4Te7}


\end{document}